\documentclass[12pt,preprint]{aastex}

\usepackage{graphicx}
\usepackage{subfigure}

\shorttitle{Star Formation History in the SMC}
\shortauthors{No\"el, N.E.D. et al.}

\begin{document}


\title{OLD MAIN-SEQUENCE TURNOFF PHOTOMETRY IN THE SMALL MAGELLANIC CLOUD. II. STAR FORMATION HISTORY AND ITS SPATIAL GRADIENTS}

\author{Noelia E. D. No\"el}
\affil{Instituto de Astrof\'\i sica de Canarias. 38200 La
Laguna. Tenerife, Canary Islands, Spain.}
\email{noelia@roe.ac.uk}

\author{Antonio Aparicio}
\affil{Instituto de Astrof\'\i sica de Canarias. 38200 La
Laguna. Tenerife, Canary Islands, Spain.}

\author{Carme Gallart}
\affil{Instituto de Astrof\'\i sica de Canarias. 38200 La
Laguna. Tenerife, Canary Islands, Spain.}

\author{Sebasti\'an L. Hidalgo}
\affil{Instituto de Astrof\'\i sica de Canarias. 38200 La
Laguna. Tenerife, Canary Islands, Spain.}

\author{Edgardo Costa}
\affil{Universidad de Chile, Departamento de Astronom\'\i a, Casilla 36-D, Santiago, Chile.}
  \and

\author{Ren\'e A. M\'endez}
\affil{Universidad de Chile, Departamento de Astronom\'\i a, Casilla 36-D, Santiago, Chile.}







\begin{abstract}

We present a quantitative analysis of the star formation history  (SFH) of
12 fields in the Small Magellanic Cloud (SMC) based on  unprecedented deep
[(B-R),R] color-magnitude diagrams (CMDs).  Our fields reach  down to the
oldest main sequence (MS) turnoff with high photometric  accuracy, which is
vital for obtaining accurate SFHs, particularly at intermediate and old ages.  We use the IAC-pop code  to obtain the SFH, using a single CMD generated using IAC-star. We obtain the  SFH as a
function  $\psi(t,z)$ of age and metallicity. We also consider several
auxiliary functions:  The Initial Mass Function (IMF), $\phi(m)$,  and a
function accounting for the frequency and relative mass distribution of
binary stars, $\beta(f,q)$.   We find that there are four main periods of
enhancement of star formation: a young one peaked at $\sim$0.2-0.5 Gyr
old, only present in the eastern and in the central-most fields; two at
intermediate ages present in all fields (a conspicuous one peaked at $\sim$4-5 Gyr old, and a  less significant one peaked at $\sim$1.5-2.5); and an old one, peaked at $\sim$10 Gyr in all fields but the
western ones. In the western fields, this old enhancement splits into two, one peaked at
$\sim$8 Gyr old and another at $\sim$12 Gyr old. This ``two-enhancement'' zone
seems to be a robust feature since it is unaffected by our choice of 
stellar evolutionary library but more data covering other fields of the SMC are necessary in order
to ascertain its significancy. 

Correlation between the star formation rate enhancements and SMC-Milky Way encounters
is not clear. Some correlation could exist with encounters taken from the orbit determination 
of Kallivayalil, van der Marel, \& Alcock (2006). But our results would be also fit in a 
first pericenter passage scenario like the one claimed by Besla et al. (2007). 
For SMC-Large Magellanic Cloud (LMC) encounters, we find a correlation only for the most recent
encounter $\sim$0.2 Gyr ago. This coincides with the youngest  $\psi(t)$ enhancement
peaked at these ages in our eastern fields.

The population younger than 1 Gyr old in the wing  area  represents  
$\sim$7-12\%  of the total $\psi(t)$. This does not reflect an exceptional
increment in the present star formation as compared with the  average
$\psi(t)$ but it is very significant in the sense that these eastern fields
are the only ones of this study in which star formation is currently going on. There is a
strong dichotomy between East/Southeast and West in the current irregular
shape of the SMC.   We find that this dichotomy is produced by the youngest population and
began $\sim$1.0 Gyr ago or later. 

The age of the old population is similar at all
radii and at all azimuth and we constrain the age of this oldest population
to be older than $\sim$12 Gyr old. We do not find yet a region dominated
by an old, Milky Way-like, halo at 4.5 kpc from the SMC center, indicating
either that this old stellar halo does not exist in the SMC or that its
contribution to the stellar populations, at the galactocentric distances of
our outermost field, is negligible. Finally, we  derive the age-metallicity
relation and find that, in all fields, the metallicity increased
continuously from early epochs until the present. This is in good agreement
with the results from the CaII triplet, a completely independent method,
constituting external consistency proof of IAC-pop in determining the
chemical enrichment law. 

\end{abstract}




\section{INTRODUCTION} \label{intro}


The Local Group dwarf galaxies provide a unique laboratory for studying and
testing galaxy formation theories and cosmology.  Their close proximity
allows individual stars to be resolved, giving accurate kinematics (see
e.g. Walker et al. 2006), photometry (see e.g. No\"el et al. 2007,
hereafter Paper I) and spectroscopy (e.g. Carrera et al. 2008a). Their
stellar populations can be characterized in detail and their star formation
histories (SFHs) derived (e.g. Gallart et al. 1999). Their extended edges
can be compared with cosmological predictions to give useful constraints
(e.g. No\"el \& Gallart 2007); and their large mass-to-light ratios can be
used, through dynamical modelling, to place constraints on the nature of
dark matter (e.g. Kleyna et al. 2001).


Containing stars born over the whole lifetime of a galaxy, the color
magnitude diagram (CMD) is a fossil record of the SFH.   For the Milky Way
satellites, it is possible to obtain accurate SFHs, from CMDs reaching the
oldest main-sequence (MS) turnoffs,  using ground-based telescopes. 
Reaching the oldest MS turnoffs is vital for breaking the age-metallicity
degeneracy and properly characterising the intermediate-age and old 
population (see Gallart, Zoccali, \& Aparicio 2005).  The Magellanic Clouds
(MCs), our nearest irregular satellites, provide an ideal environment for
this work. In this paper, we focus on the Small Magellanic Cloud
(SMC). The SMC has been historically neglected in favor of its larger
neighbor, the Large Magellanic Cloud (LMC). However, recently there has
been growing interest in the SMC as a result of new proper motion
measurements --which constrain the past orbital motions of the
MCs  (Kallivayalil et al. 2006; Piatek et al. 2008; Costa et al. 2009).
 These indicate that it may
have a different origin  to the LMC (see e.g. Bekki et al. 2004).  If
true, this would imply that its SFH, evolution and structure could differ
significantly from that of the LMC. 

The SMC lies at a distance of 61.1\,kpc from the sun  (Westerlund 1997;
Storm et al. 2004; Hilditch, Howarth, \& Harries 2005; Keller \& Wood
2006), has a  mass interior to 3 kpc of 
M$_{SMC}$$\thicksim$3$\times$10$^{9}$M$_{\odot}$ (Harris \& Zaritsky 2006),
a high fraction of HI  (M$_{HI}$$\thicksim$4$\times$10$^{8}$M$_{\odot}$,
Stanimirovi$\check{c}$ et al. 1999), a luminosity of 6$\times$10$^{8}$
L$_{\odot}$ in the {\it V}-band (de Vaucouleurs et al. 1991), and a current
metallicity of $\thicksim$1/5 solar (Dufour 1975; Peimbert \&
Torres-Peimbert 1976; Dufour \& Harlow 1977; Peimbert, Peimbert, \& Ruiz
2000). The SMC is actively forming stars at  a global rate of 
0.05M$_{\odot}$/yr (Wilke et al. 2004), and is populated by well-studied
HII regions and star clusters of all ages (e.g. Massey 2002; Rafelski \&
Zaritsky 2005; Chiosi et al. 2006; Bica et al. 2008; Piatti et al. 2008;
Glatt et al. 2008b).


\subsection{The SMC stellar content from field stars}

The most comprehensive study of the SFH of the SMC to date was presented by
Harris \& Zaritsky (2004; hereafter HZ04)\footnote{Many other recent
studies have also made valuable contributions. For example, Cioni et al.
(2006) compared the {\it k} magnitude distribution of the SMC  asymptotic
giant branch stars obtained from DENIS and 2MASS data with theoretical
distributions. They found that the SMC is on average 7-8 Gyr old,  but that
there are older stars present at its periphery while younger stars are
located towards the LMC.}.  They derived the global SFH of the SMC, based
on the Magellanic Clouds Photometric Survey (MCPS; Zaritsky et al. 1997)
{\it UBVI}  catalog that includes over 6 million SMC stars. They used the 
StarFISH package (Harris \& Zaritsky 2001) to determine the global SFH of
the  SMC, derived by summing the star formation rate (SFR)  over all 351
subregions and using three different metallicities.  They found that there
was a significant epoch of star formation up to 8.4 Gyr ago when
$\thicksim$50\% of the stars were formed,  followed by a long quiescent
period in the range 3 Gyr$\leq$age$\leq$8.4 Gyr,  and a more or less
continuous period of star formation starting 3 Gyr ago and extending to the
present.  They also found three peaks in the SFR, at 2-3 Gyr, at 400 Myr,
and 60 Myr ago. 

While global studies of the SMC like HZ04 are invaluable in aiding our
understanding of the evolution of the SMC, their CMDs do not go deep enough
to derive the full SFH from the information on the MS (B$\sim$22,
corresponding to stars  younger than $\sim$3 Gyr old on the main sequence).
Obtaining CMDs reaching the oldest MS turnoff is essential in order to
properly constrain the intermediate-age and old population (e.g. see Paper
I; Gallart, Zoccali, \& Aparicio 2005, for a review). Going deep usually
means sacrificing the available field of view so such studies are very
complementary to galaxy-wide surveys like HZ04. To our knowledge, the
papers which have presented CMDs reaching the oldest MS turnoffs so far,
studying a small field of view are: Dolphin et al. (2001), McCumber et al.
(2005), and  Chiosi \& Vallenari (2007). Dolphin et al. (2001) presented a
combination of HST and ground-based {\it V} and {\it I} images of a SMC
field situated 2${\degr}$ northeast of NGC 121. Using the  ground-based CMD
(for statistical reasons), with the Girardi et al. (2000)  models, they 
quantitatively determined the SFH for that field and found a broadly peaked
SFH, with the largest star formation rate occuring between 5 and 8 Gyr ago,
and some small amount of star formation going on since a very early epoch
and down to $\simeq$ 2 Gyr ago. McCumber et al. (2005) analyzed the stellar
populations of a SMC field located in the wing area with observations from
the HST WFPC2.  They compared the luminosity function from their observed
CMD with  those obtained from two different model CMDs, one with constant
$\psi(t)$ and another with bursts of star formation at $\sim$2 and  at
$\sim$8 Gyr. They found that the population appears to have formed largely
in a quasi-continuous mode, with a main period of star  formation between 4
and 12 Gyr ago and a very prominent recent star formation event producing
bright stars as young as 100$\pm$10 Myr.  Using deep CMDs obtained with the
ACS,  Chiosi \& Vallenari (2007) retrieved the SFH of three  fields around
SMC clusters.  The fields are located at galactocentric distances of
$\sim$0.22 kpc and  $\sim$0.45 kpc toward the East, and at $\sim$0.9 kpc 
in the southern direction.   Chiosi \& Vallenari found two main episodes in
the SFR, at 300-400 Myr and between 3 Gyr and 6 Gyr. They also found that
the SFR was low until $\sim$6 Gyr ago, when few stars were formed.

\subsection{The Stellar Populations of the outer reaches of the SMC}
 
Photometric studies of the outer SMC began  with the pioneering work  of
Gardiner \& Hatzidimitriou (1992).  With a rather shallow photometry
(reaching the horizontal branch (HB) level at R$\sim$20 mag), they mainly
gave information  about the young populations (age $\leq$2 Gyr). From their
CMDs  and contour plots of the surface distribution of MS stars with {\it
B-R}$<$0.1 and {\it R}$<$20, they noticed the almost complete absence of
bright MS stars in the northwestern part, while a considerable bright  MS
population was present in the eastern and southern area.  With the aid of
luminosity functions they found that young populations ($<$0.6 Gyr in age)
are concentrated towards the center of the SMC and in the
``wing''\footnote{The wing is located in the eastern side of the SMC,
facing the LMC. } region. Using an index defined as the difference between
the  median color [in (B-R)] of the red clump (RC) and the color of the red
giant branch (RGB) at the level of the HB, the authors inferred that the
bulk of the field population  has a median age around 10-12 Gyr. 

More recently,  Harris (2007) presented the SFH of the young inter-Cloud
population along the ridgeline of the HI gas that forms the Magellanic
Bridge and found an intermediate-age and old population at  4.4$\degr$ and
4.9$\degr$ from the SMC center in that direction, but only a young
population belonging to the SMC at 6.4$\degr$ ($\thicksim$7.2 kpc). At the
same time, No\"el \& Gallart (2007), presented the analysis of  three SMC
fields, located in the southern outskirts of the SMC. They found the first
evidence of intermediate-age and old stars belonging to the SMC at
5.8$\degr$ (6.5 Kpc) from the SMC  center. These studies together suggest
that the SMC is more extended than previously thought. 

 \subsection{Context of the present work} 

In Paper I, we presented the isochrones and color functions analysis of
twelve unprecedented deep {\it BR}-based SMC CMDs corresponding to fields
ranging from $\sim$1{\degr} ($\thicksim$1.1 kpc) to $\sim$4{\degr}
($\thicksim$4.5 kpc) from the SMC center. The fields are distributed in
different parts of the SMC, avoiding the central area (see
figure~\ref{SMC}). Each field reaches down to the old MS turnoffs, allowing
for a good characterisation of the  intermediate-age and old population in
these areas. The western fields contain very few stars younger than
$\thicksim$3 Gyr, while the fields located towards the east --the wing
region-- show very active current star formation.  The presence of
considerable amounts of young population in the eastern fields and lack
thereof in the western ones is in good correspondence with the existence or
absence of large amounts of HI at the corresponding locations
(Stanimirovi$\check{c}$ et al. 1999).  A significant intermediate-age
population is present in all of our fields.
 
In this paper, we extend the analysis presented in Paper I and obtain
quantitative SFHs of all the analyzed fields using the
IAC-pop code (Aparicio \& Hidalgo 2009). IAC-pop allows us to compare the observed CMD with synthetic CMDs generated using IAC-star (Aparicio \& Gallart 2004).  To compute the
synthetic CMDs, suitable stellar evolution libraries and ingredientes were
adopted.

The SFH of the SMC as derived from CMDs that reach the oldest MS turnoffs allows us to address several important questions, also posed in paper I: (i) What is the
age distribution of the old and intermediate-age population?; (ii) Are there
gradients in the composition of this underlying population?; and (iii) Shallower
studies inform us about the young population, but does this young
population reflect an exceptional increase of the star formation at the
present time with respect to the average SFR?

This paper is organized as follows. In \S~\ref{Data}, we briefly summarize
the characteristics of the SMC data. In \S~\ref{SFH_general}, we explain
the procedure we followed to quantitatively retrieve the SFH. In
\S~\ref{Ingredients}, we discuss the ingredients of our models, such as the
input stellar evolution models, the IMF, the characteristics of the binary
star population, and the parameterization of the SFH, among others. In
\S~\ref{SFH_SMC}, we present the detailed SFH of our  SMC fields. Finally,
in \S~\ref{discu_conclu} we discuss our results and present our
conclusions.

\section{The SMC data} \label{Data} 
 
{\it B} and {\it R} band images of twelve
8.85$\arcmin$$\times$8.85$\arcmin$ SMC fields were obtained throughout a
four year campaign (2001-2004) using the 100-inch telescope at Las Campanas
Observatory (see figure~\ref{SMC}). Photometry of the stars in all the SMC
fields was obtained using the set of DAOPHOT, ALLSTAR, and ALLFRAME
programs (Stetson 1994) and the final photometry was calibrated to the
Johnson-Cousins system. A total of 215,121 stars down to ${\it R}$$\sim$$24$
were kept, with small photometric errors  ($\sigma$$\leq$$0.15$,
$CHI$$\leq$$2.5$, and $-0.6$$\leq$$SHARP$$\leq$$0.6$).  See Paper I for a
complete description of the data reduction and photometry. 

\section{Deriving the SFH of a system} \label{SFH_general} 

The first step in accurately determining the SFH of a system is a deep CMD
reaching the oldest MS turnoffs. The advantage of reaching the oldest MS
turnoffs is twofold: (i) stellar evolution models are more accurate along the MS than for 
for more advanced stellar evolutionary phases such as the RGB or the HB where
the corresponding physics is more complicated or uncertain; and
(ii) stars are less densely packed on the MS than in the RGB or HB where stars of very different ages are
packed together in the CMD in a small interval of color and/or magnitude,
and suffer from important age-metallicity degeneracies. The SFH is composed
of several pieces of information. We adopt here the approach of Aparicio \& Hidalgo (2009), which can be sketched as follows: since time and metallicity are the most important variables in
the problem, we define the SFH as a  function $\psi(t,z)$ such that
$\psi(t,z){\rm d}t{\rm d}z$ is the number of stars formed at time $t'$ in
the  interval $t<t'\leq t+{\rm d}t$ and with metallicity $z'$ in the
interval $z<z'\leq z+{\rm d}z$.  Where necessary, the function $\psi(t)$
--defined as an integral over metallicity of $\psi(t,z)$-- and the function
$\psi(z)$ --defined as an integral over time of $\psi(t,z)$-- will be used
to represent the time-dependent SFH and metallicity-dependent SFH,
respectively. There are also several other functions and parameters related to
the SFH, that we will consider here as auxiliary: the Initial Mass Function
(IMF), $\phi(m)$; and a function accounting  for the frequency, {\it f},
and relative mass distribution, {\it q}, of binary stars, $\beta(f,q)$, are
the main ones. Our results are not sensitive to our assumptions for $\phi(m)$ and $\beta(f,q)$. Other parameters affecting the solution of $\psi(t,z)$ are the distance and reddening  (including
differential reddening) adopted. For a detailed discussion, see  Aparicio
\& Hidalgo (2009) and Hidalgo et al. (2009).

An important limitation on the information that can be retrieved from the
empirical data,  is produced by observational effects. These include
all the factors affecting and distorting the CMD, namely the
signal-to-noise limitations, the defects of the detector and the crowding
and blending between stars. The consequences are a loss of stars, changes in
measured stellar colors and magnitudes,  and external errors, which are
usually larger and more difficult to control than internal ones (Aparicio
\& Gallart 1995). A realistic simulation of observational effects is
necesary in order to obtain an accurate solution for $\psi(t,z)$.  In our
case, the simulation of the observational effects in the synthetic CMDs was
performed on a star-by-star basis,  using an empirical approach that makes
no assumption about the nature of the errors or about their propagation (Aparicio \& Gallart 1995).  Once the errors in the synthetic CMD are simulated, we call it {\it model} CMD. The process is fully
described in Gallart et al. (1999, and references therein) and in Paper I.

The procedure followed to find the SFH is similar to that described in
Hidalgo et al. (2009). The SFH is derived through a comparison
of the distribution of stars in the observed CMD with that of a model CMD,
using the IAC-pop code (Aparicio \& Hidalgo 2009).  A single {\it global}
synthetic CMD was generated using the IAC-star code for each set of input
parameters. This  global synthetic CMD comprises 10$^{6}$ stars with ages
and metallicities uniformly distributed over the full interval of variation
of $\psi(t,z)$ in time and metallicity. This represents a constant SFR as a
function of time with  equally probable metallicity, within a given range,
for each age (see ~\S~\ref{Ingredients}). Observational effects were
simulated in the global synthetic CMD as mentioned above. The  synthetic
stars were distributed in an array of {\it partial models}, $\psi_{i}$ , 
each containing stars within small intervals of age and metallicity.  Then,
a set of boxes was defined in the CMDs. In practice, two approaches may be
used: an uniform grid and an ``{\it $\grave{a}$ la carte}'' grid (see
~\S~\ref{Ingredients}).   An array, $M_i^j$, containing the number of stars
from partial model $i$ populating box $j$ is computed.  The same operation
is made in the observed CMD, producing a vector, $O^j$, containing the
number of observed stars in box $j$. This step defines the parameterization
of the CMD. 

 Any SFH (with the restriction in time and metallicity resolution imposed by the partial models) can be written as:

\begin{equation}
\psi(t,z)=A\sum_i\alpha_i\psi_i
\end{equation} 

 where $\alpha_i$$\geqslant$0 and $A$ is a scaling constant.
 The asociated distribution of stars in the defined boxes is
 
\begin{equation}
M^j=A\sum_i\alpha_iM_i^j
\end{equation} 

$M_i$ can now be compared with $O^j$ using a merit function. A reduced
Mighell $\chi^2$ (Mighell 1999), $\chi^2_\nu=\chi^2/\nu$ is used, where
$\nu=k-1$ is the number of  degrees of freedom, and $k$ is the number of
boxes used to parameterize the CMD. Minimization of $\chi^2_\nu$ with
respect to the $\alpha_i$ coeficients provides the best solution as well as
a test on whether it is acceptable, and a way to estimate errors for the
solution. IAC-pop makes use of  a genetic algorithm for the minimization of
$\chi^2_\nu$. Considering the large number of dimensions of the problem
($n\times m$, n age intervals and m metallicity intervals), such an
efficient solving procedure is required.

\section{Retrieving the SFHs for the SMC fields} \label{Ingredients}

We used IAC-pop (Aparicio \& Hidalgo 2009) to obtain the SFH, $\psi(t,z)$,
in our SMC fields. For the stellar evolution libraries, we used  the
overshooting  BaSTI\footnote{While finishing the present paper, the BaSTI
group found that the  stellar models for masses between 1.1M$_{\odot}$ and
2.5M$_{\odot}$ (i.e. age range $\simeq$ 1.0-4.0 Gyr) were calculated using
an outdated version of the code.  The SFHs presented here (as well as the
calculations  based on them) were obtained using the new version of BaSTI 
(see Appendix~\ref{appendix}).} (Pietrinferni et al. 2004; 2006; see also
Cordier et al. 2007) and Padua (summarized in Bertelli et al. 1994). 
Bolometric corrections from Castelli \& Kurucz (2003) were adopted. The
input SFR, $\psi(t)$, was chosen to be constant  between  13 Gyr
ago\footnote{The results from the WMAP (Spergel et al. 2003)  imply that
the age of the universe is 13.7$\pm$0.2 Gyr.  The first stars started
forming $\thicksim$0.4 Gyr after the beginning of the Universe. With the
current most commonly accepted distance scale for globular clusters
(Carretta et al. 2000), the age of the oldest globular cluster in the Milky
Way, derived using up to date stellar evolution models, is in good
agreement with the age of the Universe.} and now.  Kroupa's revised
IMF\footnote{m$^{-1.3}$ for 0.1$\leqslant$m/M$_{\odot}$$<$0.5 and
m$^{-2.3}$  for 0.5$\leqslant$m/M$_{\odot}$$<$100.}, $\phi(m)$, was used
(Kroupa et al. 2003).  We assumed a low metallicity bound $z_i=0.0001$,
since it is compatible with the CMD and is the lowest metallicity allowed by
the models. The high metallicity bound was taken from the HII region
observations (Dufour 1984)  (see  below and
table~\ref{metallicity_intervals}).  It is not possible to uniquely
determine the binary fraction, but we explored the consequences of the
presence of binaries  with properties similar to those observed locally on
the CMDs of the SMC. Only in binaries with mass ratios {\it q} close to
unity would the secondary have a substantial effect on the combined
luminosity of the binary. For this reason, in our final models we have 
considered mass ratios in the interval 0.7$\lesssim${\it q}$\lesssim$1.0
(see Gallart et al. 1999 for details). After testing different binary
fractions, we found that, in general, the $\psi(t,z)$ is not  significantly
affected by changes in {\it $\beta(f,q)$}. In our final models we
considered a 30\% of binary fraction.    Finally, in order to obtain the
global model CMD, we simulated the observational errors as mentioned in
~\S~\ref{SFH_general}.  We used a distance modulus of $(m-M)_{0}=18.9$ and
the reddening values given in table~\ref{reddening} (see Paper I for
details on the reddening determinations).  

 Each model CMD was divided into partial models, using the age-metallicity
pairs defined in  tables~\ref{age_intervals},~\ref{metallicity_intervals},
and~\ref{PAIRS}. In table~\ref{age_intervals}, the name of each set of age
intervals is shown in the first column and the sampling of such intervals 
is presented in the second column.  Three different set of age intervals
were used in order to address how the SFH is affected by changes in such
age intervals.  In table~\ref{metallicity_intervals}, the name of each set
of metallicity intervals is shown in the first column and the sampling of
such intervals  is presented in the second column.  The two different sets
of  metallicity intervals were chosen according to the stellar population
present in each field. In those fields in which there is a
considerable amount of young stars (eastern fields and the two closest
southern ones), the metal-1 set of intervals from table~\ref{metallicity_intervals}
was used (which reaches higher metallicities), while in those in which the
recent star formation is negligible, metal-2 from
table~\ref{metallicity_intervals} was used. Table~\ref{PAIRS}  defines the
combination of intervals of age and metallicity used for each field. The
first column gives  the number of simple populations; the second and third
columns denote the number of age and metallicity intervals, respectively; and in the
fourth column, the   corresponding  fields are shown. The age intervals are
defined such that they are larger towards older ages. This is because older
stars are more densely packed in the CMD and the isochrones become closer
together as they get older, while stars have higher photometric errors at
fainter magnitudes. By choosing these intervals of age for the partial
models, we are introducing an upper limit to the resolution in age of
$\psi(t)$.

The next step was the parameterization of the data. 
Instead of using an uniform grid, it is better to use one in which the box is different across the
CMD. We call this 
``{\it $\grave{a}$ la carte}''  parameterization (see Hidalgo et al. 2009). 
In this way, regions for which stellar positions as function of mass, age and metallicity, 
as provided by the stellar evolution theory, are better known, are sampled with smaller boxes, 
so receiving a larger weight in the solution searching. 
  We
performed several tests using different  ``{\it $\grave{a}$ la carte}''
parameterizations. Figure~\ref{boxes} shows some examples of the
parameterizations we performed and their corresponding solutions for
$\psi(t,z)$ for field smc0057. As seen from the figures, the different 
SFHs are very similar and the resulting $\chi^{2}_{\nu,min}$ are very good
in all cases, implying that the parameterization is not significatively
affecting the solution.  We kept the  ``{\it $\grave{a}$ la carte}''
parameterization shown in figure~\ref{carte}, which has small boxes in the
regions in which the stellar evolutionary phases are well known (MS), and
larger boxes in the regions of the CMD in which  stars in more advanced
phases  are located. The solution for the $\psi(t,z)$ in field smc0057 is
shown in figure~\ref{SFH_BASTI_carte}.

For all fields,  we retrieved the SFH using both stellar evolution
libraries as  inputs of IAC-star: BaSTI and  Padua. The results are  
presented in section~\ref{SFH_SMC}.

\subsection{Testing the pipeline: recovering the SFH of  ``mock'' galaxies} \label{mock}
 
Several tests of IAC-pop are discussed by Aparicio \& Hidalgo (2009) and by
Hidalgo et al. (2009). We have performed some more tests for our particular
case,  setting out to recover the SFH of two ``mock'' galaxies, generated
using the IAC-star code.  One mock galaxy assumed a constant $\psi(t)$=1 
and a metallicity law $\psi(z)$ suitable for our SMC fields (the
``SMC-mock"; see below and Carrera et al. 2008b). The other assumed the
same $\psi(t)$ but a different $\psi(z)$,  in order to investigate if the
assumption of a given metallicity law affects  the results (the
``metal-mock"). In both cases, 500,000 stars were considered. We  simulated
observational errors for each  synthetic population as described in
\S~\ref{SFH_general}.  Errors from the observed  field qj0116 were
simulated since it is a typical ``wing'' field, with a fairly large amount
of stars.  The same test was performed simulating observational errors from
other fields, obtaining similar results. 

Different subsamples were extracted randomly from SMC-mock. In each case,
$\psi(t)$ was recovered using a global synthetic CMD with 10$^{6}$ stars
computed assuming exactly the same inputs of binariety, IMF, stellar
evolution library, and bolometric corrections  as the SMC-mock.  The
metallicity distribution was assumed equally probable between $z=0.0001$
and $z=0.02$,   for the whole age interval. The resultant $\psi(t)$ are
displayed in figure~\ref{tests_normalization}. Note that  they
deviate from the input $\psi(t)$=1 by up to 25\%, showing ``wiggles'' with
similar patterns in the different subsamples.  This shows that the effect
is a {\it systematic} error, rather than a random one. It is worth noting that this effect is not caused by the crowding  present in the
different fields, as shown in figure~\ref{mock_SFH}. This figure shows the
SFH derived for  SMC-mock (for the age interval age-1 from
table~\ref{age_intervals}) obtained after simulating the observational
effects from three fields located at different galactocentric distances:
the central-most field, smc0057 (located at $\thicksim$1.1 kpc),  the
outer-most one (at $\thicksim$4.5 kpc) and field smc0049, located at an
intermediate distance from the SMC center (at $\thicksim$3.3 kpc). 

Since it is a systematic error, it should be corrected. For such purpose,  
every solution obtained with this global model CMD should be divided by
the  solution shown in figure~\ref{tests_normalization} (SMC-mock with
500,000 stars). To test further if such a correction really improves the
solutions, we performed a new test, now using a third mock galaxy with a
$\psi(t)$ similar to the solutions found for our real SMC fields. The results
are   shown in figure~\ref{qj0116test}, in which the input $\psi(t)$ for
the  mock galaxy is represented by the solid line. The recovered SFH for
such mock galaxy is represented by the dashed lines in the figure. The
recovered $\psi(t)$ differs from the input $\psi(t)$ in the same locations
as seen in figure~\ref{tests_normalization}: the recovered $\psi(t)$ is
higher than the input one at $\thicksim$8 Gyr old and lower at
$\thicksim$10 Gyr old.  The obtained $\psi(t)$ (after dividing) is in
excellent agreement with the input $\psi(t)$ as seen from the figure
(dotted line). 

Given that the  observational errors differ from field to field,  we
recovered the  systematic signature on the SFH using the SMC-mock SFH for
each of the SMC fields,  and for the three different sets of age
intervals.   Then, we divided each SFH we obtained by the corresponding
systematic signature.

\section{The SFH of the SMC fields} \label{SFH_SMC}

In order to reduce sampling problems associated
with age binning, we obtained three  different solutions for the SFH,
$\psi(t,z)$ (see Aparicio \& Hidalgo 2009) of each field, using 
three different age-binning sets (see
table~\ref{age_intervals}).
The adopted solution will be the average of the three. As an example, 
figure 7 shows the three solutions obtained with the BaSTI library for field smc0057
together with the adopted solution, the age-metallicity relation and the observed CMD. 
 The left panel shows the 3D population boxes (Hodge 1989) of the three
solutions, where $\psi(t,z)$ is represented as a function of age and metallicity. 
Here, the volume of each bar over the age-metallicity plane gives
the mass that has been transformed into stars within  the  corresponding
age-metallicity interval.
The adopted solution (right medium panel) is obtained as a cubic spline fit to the three individual solutions
after correcting for the systematic errors discussed above. 
As in figure 3, error bars (vertical) are only indicative, while the actual dispersion of the three
solutions (see Aparicio \& Hidalgo 2009) should be considered a more realistic representation of the solution
uncertainties. The age-metallicity relations shown in the bottom panel have been obtained as the 
average metallicity of the stars in each age interval for the three individual solutions. 
  
The solutions obtained using the Padua library are  very similar to the
ones obtained using the BaSTI library and  are not presented in the detail
of figure~\ref{figure_3D}.  Figures~\ref{SFH_I}
and~\ref{noeplot_final_bertelli} show a summary of the  results obtained
for all fields using BaSTI and Padua, respectively. They show, for each
field, the spline fit  together with the results for the 3 age binning
sets. From now on, our discussion of the results will use the
results from the BaSTI stellar evolution library. Our conclusions
are unchanged if we use instead the results from the Padua library.

\subsection{Main characteristics of the $\psi(t)$ solutions for our SMC fields} \label{main_characteristics}  
 
As seen from figure~\ref{SFH_I},  the eastern fields and the central-most
field, smc0057, --located in the south-- show a large amount of recent star
formation. In particular,  the eastern fields show a recent enhancement
from $\thicksim$2 Gyr ago until the present, while  smc0057 shows a recent
peak of star formation  $\thicksim$1 Gyr ago, which seems to be mostly
extinguished at the present time. This is in agreement with the
characteristics derived for the stellar populations in the  Magellanic
Bridge  (Harris 2007) and in other positions in the wing area of the SMC
(see, for example,  Irwin et al. 1990; McCumber et al. 2005; Chiosi \&
Vallenari 2007; among others). These $\psi(t)$ enhancements at young ages
in the eastern fields  and in smc0057 are not seen in other fields located
at similar galactocentric distances.  The three eastern  fields --the only
ones presently forming stars-- are  located in regions of large amount of
HI, unlike  the rest of our fields, including smc0057.

A  conspicuous intermediate-age enhancement peakes between $\thicksim$4 and
$\thicksim$5 Gyr old  in all fields. In addition, there is a small
enhancement at $\thicksim$2-2.5 Gyr old in the southern and in the western
fields.   This $\psi(t)$ enhancement is shifted toward younger ages, at
$\sim$1.5-2 Gyr old in the eastern fields.  Finally, a $\psi(t)$
enhancement at old ages peaks at $\thicksim$10 Gyr old in the eastern and
southern fields, which seems to be ``split'' into two, at $\thicksim$8 and
$\thicksim$12  Gyr old, in the western fields.   
Note that most of the above features remain unchanged when using the Padua
stellar evolution library, as seen in figure~\ref{noeplot_final_bertelli}. 

\subsection{Global bursts and phase mixing in the SMC} \label{well_mixed}

Phase mixing in a galaxy occurs when stars initially close in space --for
example stars formed in a star forming  region-- spread out over time
because they have slightly different energies and angular momenta. Stars
are said to be fully phase mixed if there is no memory left that they were
born close together. The rate at which stars phase mix depends on the
gravitational potential, on the initial proximity of the stars, and on
their orbits. As a consequence of the latter,  perfectly circular orbits
will never mix in radius, while perfectly radial orbits never mix in
angle.

The presence of  the $\psi(t)$ enhancements at $\sim$4-5 Gyr old in all the SMC fields, together with 
the large variations found for ages younger than $\sim$2 Gyr old, would
suggest that the phase mixing time in the SMC is of the order of $\sim$2
Gyr. However, we find also evidences for spatial variations at older ages:
the western fields present two $\psi(t)$ enhancements at $\sim$8 Gyr old
and at $\sim$12 Gyr old, while in the rest of the fields there is a single
old enhancement  occurring $\sim$10 Gyr ago.  This could imply that stars in the SMC take a Hubble time
or more to phase mix.  
However, solutions are noisier and time resolution is worst for older ages, for which this conclusion 
must be taken very cautiously until more accurate and precise data, sampling a larger area, are
available. 

\subsection{Spatial distribution of the stellar populations in our SMC fields} \label{distribucion_espacial}

One of the most intriguing  issues regarding the SMC evolution is the age
and distribution of its oldest stars.  In order to shed light into this, we
calculated the age at the 5th percentile of $\psi(t)$ in each of our SMC
fields, i.e., the population age by which 5\% of the total stars were
formed in each field, which is also presented in figure~\ref{percentiles}. 
The 5th percentile age in all fields presents a flat distribution at
$\thicksim$11.5 Gyr. In fact,  the slope of the best-fit line in a
least-square fit is 0.064$\pm$0.015 (almost negligible).  This shows that
the age of the  old population in all our SMC fields is essentially the
same, independently of the galactocentric distance or the position angle. 
In addition, we constrain the age of the oldest population to be older than
$\thicksim$11.5 Gyr old. This is in agreement with the recent age
determination of the older and single globular cluster in the SMC, NGC121
(Glatt et al. 2008). Our results are also in good agreement with those of 
Dolphin et al. (2001) who, for an isolated field, located
in the northwestern part of the SMC, found that 14$\pm$5\% of the star
formation took place before 11 Gyr ago. 
 
 Another important --and controversial-- fact regarding Local Group dwarf
galaxies in general and the SMC in particular, is the composition of the
outer extended stellar populations.  The 95th percentile age for all the
SMC fields shows a relatively flat distribution (except for the dichotomy
East-West in the  central-most fields) while going further away from the
SMC center. This points out that, at 4.5 kpc from the SMC center, we did
not yet reach a region dominated by an old, Milky Way-like, stellar halo.
This is stressed by the fact that the 95th percentile age for our outermost
field occured at $\thicksim$3 Gyr ago.  If we would be in such halo
dominated region, the 95th percentile and the 5th percentile age should
occur at almost the same time for the outermost fields.  Our results are in
agreement with No\"el \& Gallart (2007) who found that up to $\thicksim$6.5
kpc from the SMC center, the galaxy is composed by both, intermediate-age
and old population. In summary, our results indicate that either an old,
Milky Way-like, stellar halo does not exist in the SMC or that if it
exists, its contribution to the stellar population is negligible at
$\sim$4.5 kpc from the galactic centre.

\subsection{On the possible correlation between 
  the $\psi(t)$ enhancements and the SMC-LMC/SMC-MW pericenter passages} \label{enhance_pericenter}

 In the pioneering work from Murai \& Fujimoto (1980),
the authors claimed that the existence of the Magellanic Bridge and the
inter-Clouds region are partly explained if the SMC closely approached the
LMC around 0.2  Gyr ago.  Since then, the orbits of the MCs were studied in
detail by many authors,  through numerical simulations and proper motion
studies (see  Gardiner et al. 1994; Bekki \& Chiba 2005; Kallivayalil et
al. 2006; Besla et al. 2007; among others). All models reproduce a
pericenter passage between the MCs around $\sim$0.2 Gyr ago.  Coincidently,
enhancements of star formation are found at these ages in both galaxies,
particularly in the area in which they are facing each other, i.e., the
wing area in the SMC and the West part in the LMC (see, for example,  Irwin
et al. (1990). Given the low temporal sampling of our SMC SFHs for the youngest ages, 
we cannot probe if the dichotomy East/Southeast-West actually began $\sim$0.2 Gyr ago. However, 
the steep behaviour of the $\psi(t)$ 95th percentile age shown in figure 10 indicates that the dichotomy 
appeared at an age smaller than $\sim$1 Gyr ago. This population younger than 1 Gyr old represents $\sim$7-12\%
of the total $\psi(t)$ in the wing area. This does not reflect an exceptional increment in the present star 
formation as compared with the average $\psi(t)$ but it is very significant in the sense that these eastern fields
are the only ones in which star formation is currently going on.

Besides this youngest episode, authors such as Bekki et al. (2004), Bekki \& Chiba (2005)
 and  Harris \& Zaritsky (2004)   have
claimed that the episodes of enhancement in the SFH of the SMC could be
related with pericenter passages between the SMC and the LMC and/or between
the SMC and the Milky Way, while Besla et al. (2007) have concluded that the Magellanic Clouds are
likely to be in their first pericenter passage about the Milky Way or on a highly eccentric, bound orbit. 
To explore this, the $\psi(t)$ enhancements in our SFHs are
quantified in figure~\ref{enhancements},  in which the intensity of each
$\psi(t)$ enhancement as a function of radius (\ref{enhancenew_d}), 
position angle (\ref{enhancenew_pa}), and age  for all the fields are
represented, together with the pericenter passages of the SMC with respect
to the Milky way or the LMC, as predicted by different authors (see figure
caption for details). The intensity of each $\psi(t)$ enhancement is
defined as the area under a Gaussian function fitted to the elevation in
the spline fit shown in figure 8.  

Although unclear, there may be  a correlation between the SMC-Milky Way encounters
given by  Kallivayalil et al. (2006; solid arrows) and the enhancements in
$\psi(t)$ we found at $\thicksim$2.5 Gyr ago, at $\thicksim$4.75 Gyr ago
and at $\thicksim$8 Gyr ago.  In the case of pericenter passages between
the LMC and the SMC, there only seems to be a coincidence between the most
recent encounter $\sim$0.2 Gyr ago and the youngest $\psi(t)$ enhancement
peaked at these ages in our eastern fields.  In the other cases, for the
published orbits, we see no clear correlation between the pericenter
passages and the observed  enhancements in our derived SFHs. All in all, the lack of 
a clear correlation between the computed passages and SFH could be a support to Besla et al. (2007) results
including that indicating that the SMC is in its first pericenter passage about the Milky Way.

\subsection{Comparison with other works} \label{compara}
  
Since our eastern and western fields, as well as two of our southern ones, 
overlap the regions from HZ04, we superimposed our SFHs with the ones they
obtained as seen in  figure~\ref{HZ04}. In each case the SFHs found by HZ04
are shown in dashed lines.  HZ04 used the starFISH code with the following
inputs:  a subset of the Padua isochrones for three different 
metallicities (Z=0.001, Z=0.004, Z=0.008) without interpolation, a power
law with Salpeter slope for the IMF, and a 50\% binary fraction with
secondary masses drawn randomly from the IMF.  We averaged the SFR from
HZ04 in the last 0.2 Gyr  into only one age bin  to fit the age resolution
we adopted for the youngest population. We also added up the SFR given by
HZ04 for each of their three metallicities. HZ04 cover a larger area in the
region of our western fields qj0036 and qj0037, so the solution they give 
includes both of our fields.   We compare the HZ04 solution with ours obtained
using the Padua stellar evolution library as input in IAC-star/IAC-pop
(shown in \S~\ref{SFH_SMC}). With the exception of field qj0033, there is a general disagreement at
intermediate ages in all fields, for
which HZ04 find either total or quasi-total quiescence during $\thicksim$2
Gyr of the life-time of the galaxy.  In the western fields, HZ04 find a
peak at around 4.5 Gyr ago which is coincident with the one we find.

In order to understand the disagreement between  our SFHs and the ones
obtained by HZ04, it should be noted that their photometry is  shallower
and, therefore, the ability to reliably constrain the intermediate-age to
old star formation is reduced. Also,  their method to derive the
SFH is coarser than the one used in this paper (for example, no
interpolation in metallicity is performed  and so the simple populations
are restricted to the metallicities provided in the stellar evolution
set).

 Our SFHs solutions for the western fields agree quite well with the SFH
presented by Dolphin et al. (2001)  (using Girardi et al. 2000 models)
for a northwestern field located near NGC121.  They find a broadly peaked
star formation between 5 and 8 Gyr ago and that the star formation almost 
stopped around 2 Gyr ago.  

The $\psi(t)$ enhancement peaked at $\sim$4-5 Gyr old is in good agreement
with the episodes found by Chiosi \& Vallenari (2007) between  3 and 6 Gyr
ago for three fields located around the SMC clusters K 29, NGC 290, and NGC
265. 

Finally, estimates by Sabbi et al. (2009) are broadly consistent with our results, although we have to 
wait for the results of the detailed, quantitative analysis of the SFH that these authors are carrying on.

 \subsection{The Chemical Enrichment History} \label{CEH}
         
In the computation of the SFH, IAC-pop also provides the age-metallicity
relation, which is plotted in the horizontal plane of
figure~\ref{figure_3D}.  To more clearly show this metallicity law, we
determined the median metallicity of stars formed at each age interval, 
using the following relation (see \S~\ref{SFH_general}): 

\begin{equation}
Z(t)=\frac{\sum  z_{i} \psi_i(t)}{\sum \psi_i(t)}
\end{equation} 

  We adopted Z$_{\odot}$=0.02 in order to convert from Z metallicities to [Fe/H] values and assuming
 [Fe/H]=log(Z/Z$_{\odot}$). The age-metallicity relation computed in this way for the eastern, western, and southern fields are
shown with solid line in figure~\ref{ceh} together with the age-metallicity relation found by 
Carrera et al. (2008b) (see their table 6) using Ca II triplet spectroscopy of RGB stars
 from the same fields analysed in
    this paper.  The $\pm$1$sigma$ dispersion of the stellar metallicity distribution as a function of age as
    derived in this paper, and the metallicity dispersion of the Ca II triplet metallicities in each bin, from 
    Carrera (2008b) have also been represented. 
 
The  age-metallicity relations in all fields (\ref{ceh_east},~\ref{ceh_west},~\ref{ceh_south}) show  a continuously 
increasing metallicity from an early epoch until now. 
   For the southern fields, 
 there is an excellent agreement with the findings of 
Carrera et al. (2008b).
The agreement is good in the case of the eastern and western fields,  with 
small differences for ages older than 5 Gyr, for which we find a lower metallicity in the west and a higher metallicity in the east than
those of Carrera et al. (2008b).
 These results, taken together, are an important test of IAC-pop because they 
show, for the first time, the external consistency of the code in determining the chemical enrichment law.  

Tsujimoto \& Bekki (2009) claim that a dip is detected in the [Fe/H]-age relation in the SMC and that 
would be related to a major merging event occurred some 7.5 Gyr ago. We have to mention that such 
dipping is not visible in the age-metallicity relations derived in our analysis.

\section{Discussion and Conclusions} \label{discu_conclu}

We have presented a detailed study of the SFH of 12 fields located in the
Small Magellanic Cloud, based on a set of [{\it (B-R), R}] CMDs that reach
the oldest MS turnoffs (M$_{R}$$\thicksim$3.5). The spatial distribution of
the fields, located at different galactocentric distances and azimuths, 
makes it  possible to distinguish the stellar content in the wing area and
in the ``undisturbed'' parts toward the western and southern regions of the
SMC (see figure~\ref{SMC}), and to study possible stellar population
variations with galactocentric radius. We used the IAC-star and IAC-pop
codes to obtain the SFH, $\psi(t,z)$. The results of this analysis allow us
to  accurately constrain the parameter space defining the SFHs of the 12
SMC fields. The fact that the main characteristics of   $\psi(t,z)$ are
unchanged for different combinations of parameters, including different
stellar evolution libraries, indicates that our solutions for  the SFHs are
robust. In addition, common patterns, which vary smootly with position,
appear in most fields. As final inputs for IAC-star/IAC-pop we used the
BaSTI  (Pietrinferni et al. 2004) and Padua (Bertelli et al. 1994) stellar
evolution libraries, the bolometric corrections from Castelli \& Kurucz
(2003), the Kroupa's revised IMF (Kroupa et al. 2003), and a 30\% of
binaries with a mass ratio ${\it q}$$\gtrsim$$0.7$. All the $\psi(t,z)$
solutions have $\chi^{2}_{\nu,min}$$<$2.

In the retrieved SFHs of our SMC fields, we found the following. There are
four main episodes of enhancement in $\psi(t)$: one at young ages, only
present in the eastern fields   (the ones facing the LMC) and in the
central-most one (located in the south), peaked at $\thicksim$0.2-0.5 Gyr
ago; two at intermediate ages, a conspicuous one peaked at $\thicksim$4-5
Gyr old in all fields and a less significant one peaked at
$\thicksim$1.5-2.5 Gyr old in all fields; and one at old ages, with the
peak at $\thicksim$10 Gyr old in all fields but the western ones, in which
this old enhancement is split into two at $\thicksim$8 Gyr old and at
$\thicksim$12 Gyr old. There are smaller enhancements and variations from
field to field that  are less significant. 

The fact that all fields present  $\psi(t)$ enhancements at
$\thicksim$1.5-2.5 Gyr old and at $\thicksim$4-5 Gyr old could mean that,
at these ages, there were  global episodes of star formation in the SMC.
Alternatively,  these episodes could have been produced in a particular
region of the SMC and then the stars could have spread all over the galaxy,
such that stars older than $\thicksim$1.5-2.5 Gyr old are well mixed, both
in radius and in azimuth. The large variations for ages younger than 
$\thicksim$1.5-2.5 Gyr old  and the common burst of $\sim$1.5-2.5 Gyr old
would suggest that the phase mixing time in the SMC is of the order of
$\sim$1.5 Gyr.   However, we find also evidence for variations at old ages
(a $\psi(t)$ enhancement at 10 Gyr old in the East and in the South but at
$\sim$8 and $\sim$12 Gyr old in the western fields). These differences at
old ages seem to be robust features. If so, they could imply that stars in the SMC 
take a Hubble time or more to phase mix. Alternatively, they could be the result of recently dissolved old star/globular clusters. In future work, it will be interesting
to determine the SFHs over larger areas at different azimuths in order to
confirm the dichotomy in the SFH at old ages and to  constrain the spatial
limits of this ``two-enhancements zone''. This, with the aid of theoretical
models, will help to address the possible origin of such enhancements in
$\psi(t)$.  

The eastern fields are located  in a region of high  HI concentration  (see
figure~\ref{SFH_I}). We found that the young population present in this
wing area in the last 1 Gyr represents between $\thicksim$7-12\% of the
total stars found in it. This indicates that, although the young population
does not reflect an exceptional increase of the star formation at the
present time with respect to the average $\psi(t)$, this increase is
important in global terms since this wing area is the only part of our study in which
there is  active and conspicuous star formation presently going on\footnote{It is worth noticing that 
the highest current star formation activity is in the central, bar region of the galaxy, which is not 
studied here. In such area, several strong HII regions are located.}. 

The young 
$\psi(t)$ enhancement 
may have been triggered by a close encounter between the 
  SMC and the LMC at these ages, as indicated by
studies of the MCs orbits, both from numerical simulations and proper
motions (Murai \& Fujimoto 1980;  Gardiner et al. 1994; Bekki \& Chiba
2005; Kallivayalil et al. 2006; among others). Given the low temporal
resolution  of our SMC SFHs for young ages, we cannot probe if
the dichotomy East/Southeast-West actually began  $\sim$0.2 Gyr ago.
However, the step behaviour of the $\psi(t)$ 95th percentile age shown in figure 10 indicates that the 
dichotomy appeared at an age smaller than $\sim$1 Gyr ago. 
 
A correlation may exists between past  $\psi(t)$ enhancements and the perigalactic encounters
between the SMC and the Milky Way for the orbits given by Kallivayalil et al. (2006). But this correlation 
is unclear and there is nothing against the Magellanic Clouds being in their first perigalactic passage as claimed
by Besla et al. (2007). On another side, with the exception of the $\psi(t)$
enhancement peaked  $\sim$0.2 Gyr ago in the Eastern fields,  we do not
find a clear correlation between the enhancements in $\psi(t)$ and the
pericenter passages between the SMC and the LMC as computed by Bekki \&
Chiba (2005) and by Kallivayalil et al. (2006).

The flat distribution at $\thicksim$12 Gyr old of the age at the 5th
percentile indicates that  the age of the oldest population is remarkably
similar in all fields at all radii and at all azimuths and constrains the
age of the oldest stars in our SMC fields to be older than 12 Gyr old. 
This is also seen in other Local Group galaxies, such as Phoenix, a smaller
and non-interacting galaxy (see Hidalgo et al. 2009).
 
We did not reach a region dominated by an old, Milky Way-like, stellar halo
at 4.5 kpc from the SMC center.  This   indicates that either an old, Milky
Way-like, stellar halo does not exist in the SMC or that if it exists, its
contribution to the stellar population is negligible at $\sim$4.5 kpc.
These results are in agreement with No\"el \& Gallart (2007) who found no
signs of an old stellar component domination at $\sim$6.5 kpc from the SMC
center.

Finally, from our SFH solutions, we also retrieved a chemical enrichment
history for our SMC fields. On average,  all fields show  a continuously 
increasing chemical enrichment from an early epoch until now.  Our derived
age-metallicity relations are in good agreement with the findings of
Carrera et al. (2008b) using the CaII triplet. This is  external
consistency proof of IAC-pop in determining the chemical enrichment law. 

\acknowledgments

We are very grateful to Mike Beasley, Ricardo Carrera, Santi Cassisi, Gary Da Costa, Ken Freeman and Matteo Monelli for interesting comments and illuminating discussions. We also thank Justin I. Read for many very useful comments and discussions which improved this manuscript in  its early stages. This work has been founded by the
Instituto de Astrof\'{\i}sica de Canarias (P3-94) and by the Ministry of
Education and Research of the Kingdom of Spain (AYA2004-06343 and
AYA2007-67913). E. C. and R. A. M. acknowledge support by the Fondo
Nacional de Investigaci\'on Cient\'{\i}fica y Tecnol\'ogica (No. 1050718,
Fondecyt) and by the Chilean Centro de Astrof\'{\i}sica FONDAP (No.
15010003).  This project has made generous use of the 10\% Chilean time,
and continuous support from the CNTAC and Las Campanas  staff is greatly
appreciated. N.E.D.N. acknowledges the hospitality of the astrophysics
group at Z\"urich University.

\begin{deluxetable}{ccccc}
\tabletypesize{\scriptsize}
\tablewidth{0pt}
\setlength{\tabcolsep}{0.1in}
\tablecaption{Reddening values}
\tablehead{
\colhead{Field} & \colhead{E(B-V)}
 }  

\startdata

smc0057 & 0.09   \\  
qj0037 &  0.07    \\  
qj0036  & 0.07   \\  
qj0111  & 0.09    \\ 
qj0112  & 0.09    \\ 
qj0116  & 0.08   \\ 
smc0100  & 0.05  \\ 
qj0047  &  0.05    \\ 
qj0033  & 0.03   \\ 
smc0049  & 0.06     \\ 
qj0102  &  0.05   \\ 
smc0053  & 0.06    \\

\enddata
\label{reddening}

\end{deluxetable}

\begin{deluxetable}{clcccc}
\tabletypesize{\scriptsize}
\tablewidth{0pt}
\setlength{\tabcolsep}{0.1in}
\tablecaption{Age intervals}
\tablehead{
\colhead{name} & \colhead{age intervals (in Gyr)}
 }  

\startdata
	 
age-1 &  0   0.5   1     2   3   4   5   7   9   11  13\\   
age-2 &  0   0.2   0.5   1.1 1.8 2.7 3.9 5.4 7.2 9   11  13\\  
age-3 &  0   0.1   0.2   0.5 1   1.9 2.7 3.3 4.1 5   6   7.1 9   10.7 13\\ 
\enddata
\label{age_intervals}

\end{deluxetable}

\begin{deluxetable}{clcccc}
\tabletypesize{\scriptsize}
\tablewidth{0pt}
\setlength{\tabcolsep}{0.1in}
\tablecaption{Metallicity intervals}
\tablehead{
\colhead{name} & \colhead{metallicity intervals}
 }  

\startdata
	 
metal-1 & 0.0001 0.0003 0.0006 0.001 0.0015 0.002 0.003 0.004 0.005  0.006 0.008 0.01 0.015 0.02  \\  
metal-2 & 0.0001 0.0003 0.0006 0.001 0.0015 0.002 0.003 0.004 0.006 0.008 \\  
\enddata
\label{metallicity_intervals}

\end{deluxetable}

\begin{deluxetable}{ccccc}
\tabletypesize{\scriptsize}
\tablewidth{0pt}
\setlength{\tabcolsep}{0.1in}
\tablecaption{Age-metallicity pairs}
\tablehead{
\colhead{Simple populations} & \colhead{age intervals} & \colhead{metallicity intervals} & \colhead{fields}
 }  

\startdata

90 & 10 & 9 &   qj0047, smc0049, qj0102, smc0053, qj0033, qj0036, qj0037  \\  
99 & 11 & 9 &   qj0047, smc0049, qj0102, smc0053, qj0033, qj0036, qj0037  \\  
126 & 14 & 9 &  qj0047, smc0049, qj0102, smc0053, qj0033, qj0036, qj0037  \\  
130 & 10 & 13 &  smc0057, smc0100, qj0111, qj0112, qj0116  \\ 
143 & 11 & 13 &  smc0057, smc0100, qj0111, qj0112, qj0116  \\ 
182 & 14 & 13 &  smc0057, smc0100, qj0111, qj0112, qj0116  \\ 

\enddata
\label{PAIRS}

\end{deluxetable}

\begin{deluxetable}{clcccc}
\tabletypesize{\scriptsize}
\tablewidth{0pt}
\setlength{\tabcolsep}{0.1in}
\tablecaption{$\chi^{2}_{\nu,min}$ for the solutions $\psi(t,z)$ in the SMC fields}
\tablehead{
\colhead{field} &  \colhead{$\chi^{2}_{\nu,min}$ age-1} & \colhead{$\chi^{2}_{\nu,min}$ age-2} &\colhead{$\chi^{2}_{\nu,min}$ age-3}
 }  

\startdata
	 
smc0057 & 1.19 & 1.36 & 1.46\\  
qj0037 & 1.12 & 1.13 & 1.26\\ 		      
qj0036  & 1.13 & 1.21 & 1.26\\ 		     
qj0111 &   1.43  &  1.45 & 1.62\\			
qj0112 &  1.41 &  1.47 & 1.57\\			
qj0116 & 1.45 & 1.57 & 1.58\\
smc0100 & 1.58 & 1.78  & 1.53\\
qj0047  & 1.64 & 1.74  & 1.78 \\
qj0033  & 0.88 & 0.93 & 0.92\\
smc0049 & 1.38 & 1.55   & 1.41\\
qj0102  & 1.33 & 1.47  & 1.31\\
smc0053 & 1.21  & 1.27  & 1.39\\
\enddata					
\label{chi2_fields}				
						
\end{deluxetable}

\clearpage
\newpage

\begin{figure}
\plotone{f1.eps}
\caption{Spatial distribution of our SMC fields. The large squares denote the 34$\arcmin$ $\times$ 33$\arcmin$ 
fields analyzed in No\"el \& Gallart (2007). The small symbols represent the fields analyzed here and in Paper I.\label{SMC}}
\end{figure}

\begin{figure*}
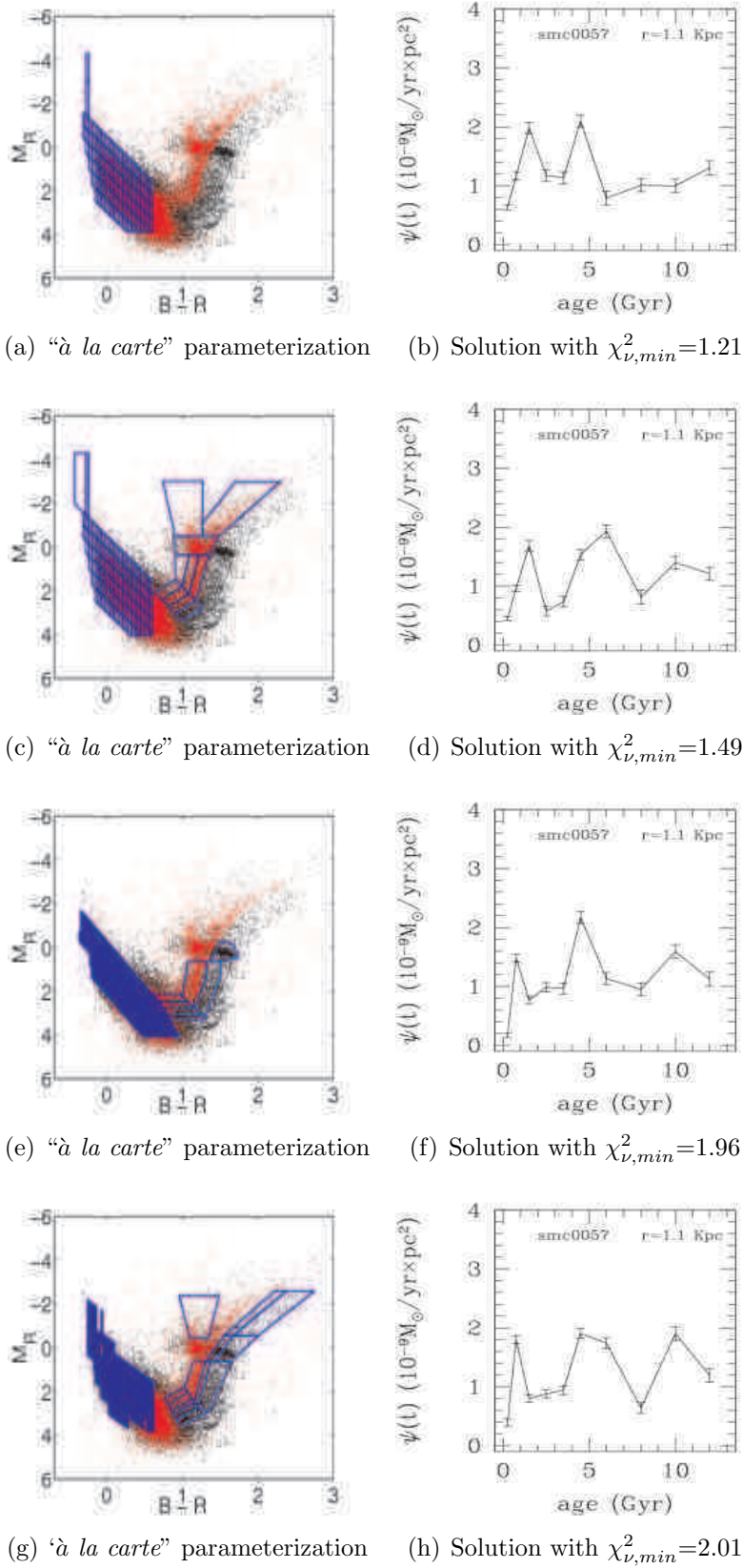

\begin{center}\subfigure[``{\it $\grave{a}$ la carte}'' parameterization]
	{
		\label{smc0057_3_modif}
\includegraphics[width=5cm,height=4.5cm]{smc0057_4_boxes.eps}
}
\subfigure[Solution with $\chi^{2}_{\nu,min}$=1.21]
	{
		\label{smc0057_4}		
\includegraphics[width=5cm,height=4.5cm]{smc0057cassisi3e1_SP.eps}
         }
	 \\
\subfigure[``{\it $\grave{a}$ la carte}'' parameterization]
	{
		\label{smc0057_3_modif}
\includegraphics[width=5cm,height=4.5cm]{smc0057_3_modif_boxes.eps}
}
\subfigure[Solution with $\chi^{2}_{\nu,min}$=1.49]
	{
		\label{smc0057_4}		
\includegraphics[width=5cm,height=4.5cm]{smc0057cassisi3e1_RC.eps}
         }
	 \\
\subfigure[``{\it $\grave{a}$ la carte}'' parameterization]
	{
		\label{smc0057cassisi3NEW1_boxes}
\includegraphics[width=5cm,height=4.5cm]{smc0057cassisi3NEW1_boxes.eps}
}
\subfigure[Solution with $\chi^{2}_{\nu,min}$=1.96]
	{
		\label{smc0057cassisiRGB}		
\includegraphics[width=5cm,height=4.5cm]{smc0057cassisiRGB.eps}
         } 	 
	 \\
\subfigure[`{\it $\grave{a}$ la carte}'' parameterization]
	{
		\label{smc0057_cassisi16_boxes}
\includegraphics[width=5cm,height=4.5cm]{smc0057_cassisi16_boxes.eps}
}
\subfigure[Solution with $\chi^{2}_{\nu,min}$=2.01]
	{
		\label{smc0057similar}		
\includegraphics[width=5cm,height=4.5cm]{smc0057similar.eps}
         } 
 
\caption{Left panels show some examples of the parameterizations we performed on the observed (red) and model (black) CMDs 
using BaSTI library. The corresponding solutions are shown in the panels on the right.
 Error bars have been computed as the dispersion of 20 solutions with  $\chi^{2}_{\nu}$=$\chi^{2}_{\nu,min}$+1, where 
 $\chi^{2}_{\nu,min}$ is the solution shown in this figure.}
\label{boxes}
\end{center}
\end{figure*}

\begin{figure*}
\begin{center}\subfigure[Using ``{\it $\grave{a}$ la carte}'' parameterization]
	{
		\label{carte}
\includegraphics[width=7.5cm,height=5.5cm]{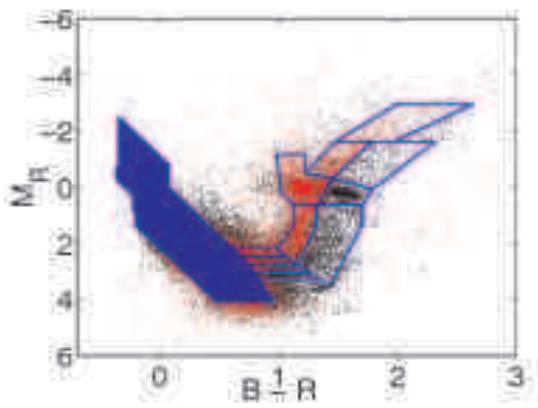}
}
\subfigure[Using ``{\it $\grave{a}$ la carte}'' parameterization]
	{
		\label{SFH_BASTI_carte}
\includegraphics[width=7.5cm,height=5.5cm]{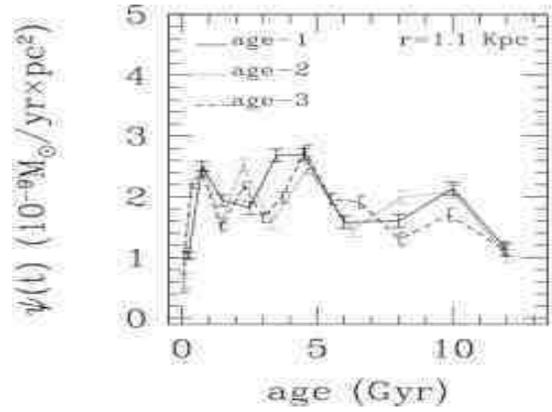}
         }
\caption{Panel~\ref{carte} shows the final set of boxes used to obtain the $\psi(t,z)$ of our SMC fields. 
  For the MS, we used a {\it quasi}-grid parameterization. We follow the isochrone's track as a guide and for the subgiant branch, the 
  RGB and the RC larger boxes were carefully selected in order to avoid introducing errors. The final solution is shown in
  \ref{SFH_BASTI_carte} for three different age binnings.}
\label{definitivo}
\end{center}
\end{figure*}

\begin{figure*}
\begin{center}
\includegraphics[width=10cm,height=10cm]{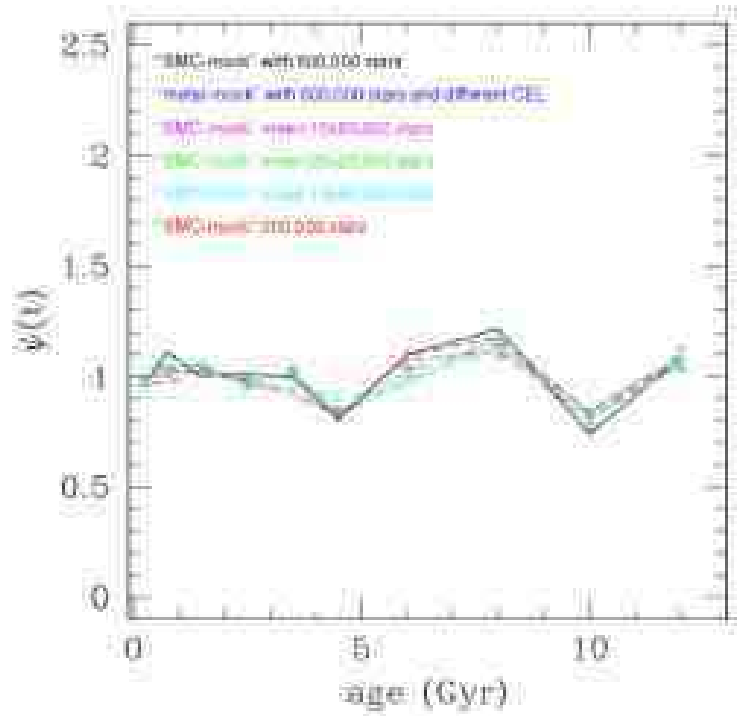}
\caption{The solution for the SFH of several samples of the ``SMC-mock'' and ``metal-mock'' synthetic populations are shown. 
The input $\psi(t)$=1 and the input metallicity law is one suitable for the SMC.  
In the case of the mean SFHs the errors are defined as: $\sigma$/(N-1)$^{1/2}$, where  $\sigma$$^{2}$ is the variance. 
 There is a systematic deviation from the input ($\psi(t)$=1) SFH, 
 showing ``wiggles''. See text for details.}
\label{tests_normalization}
\end{center}
\end{figure*}

\begin{figure*}
\begin{center}
\includegraphics[width=10cm,height=10cm]{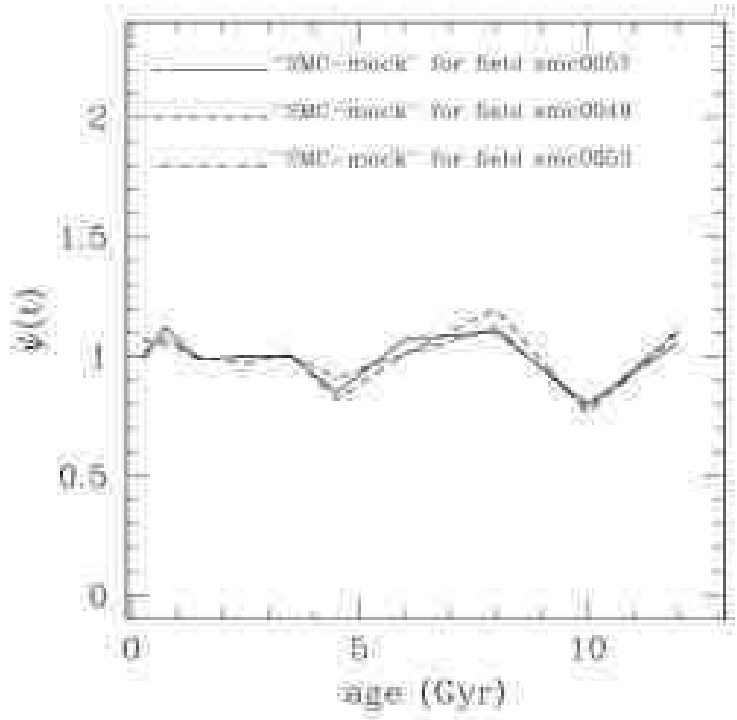}
\caption{SMC-mock SFHs obtained simulating the observational errors for
 three  SMC fields located at different galactocentric distances: smc0057 (at $\thicksim$1.1 kpc),
 smc0049 (at $\thicksim$3.3 kpc), and smc0053 (at $\thicksim$4.5 kpc). }
\label{mock_SFH}
\end{center}
\end{figure*}

\begin{figure*}
\begin{center}
\includegraphics[width=10cm,height=10cm]{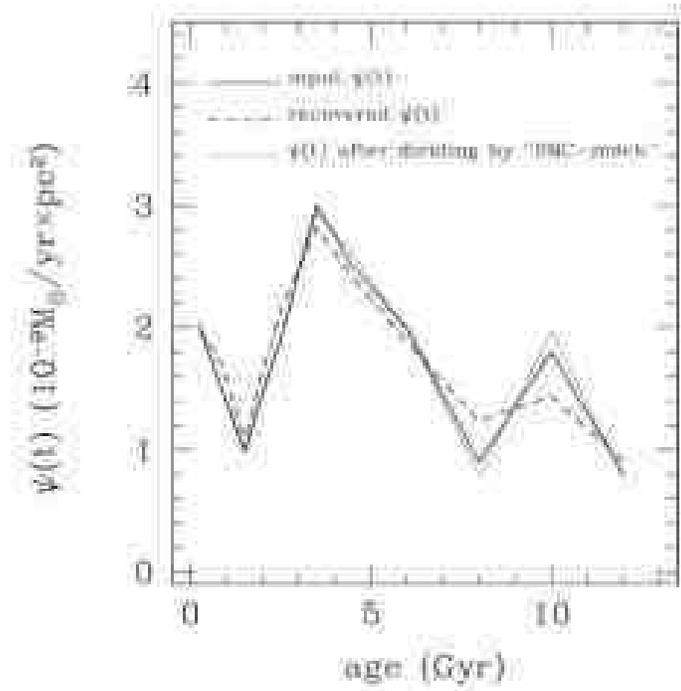}
\caption{Input, recovered, and solution $\psi(t)$ for a given galaxy with similar characteristics to the $\psi(t)$ obtained for the
 SMC fields. 
Given the input $\psi(t)$, the recovered $\psi(t)$ slightly differs in the age bins seen in
figure~\ref{tests_normalization}. It is clear that the final
 $\psi(t)$, obtained after dividing the 
input $\psi(t)$ by the ``SMC-mock'' with 500,000 stars, is in excellent agreement with the input $\psi(t)$.}
\label{qj0116test}
\end{center}
\end{figure*}

\begin{figure*}
\begin{center}
\subfigure[Field smc0057]
      {
	      \label{3D_1}
\includegraphics[width=10cm,height=15cm]{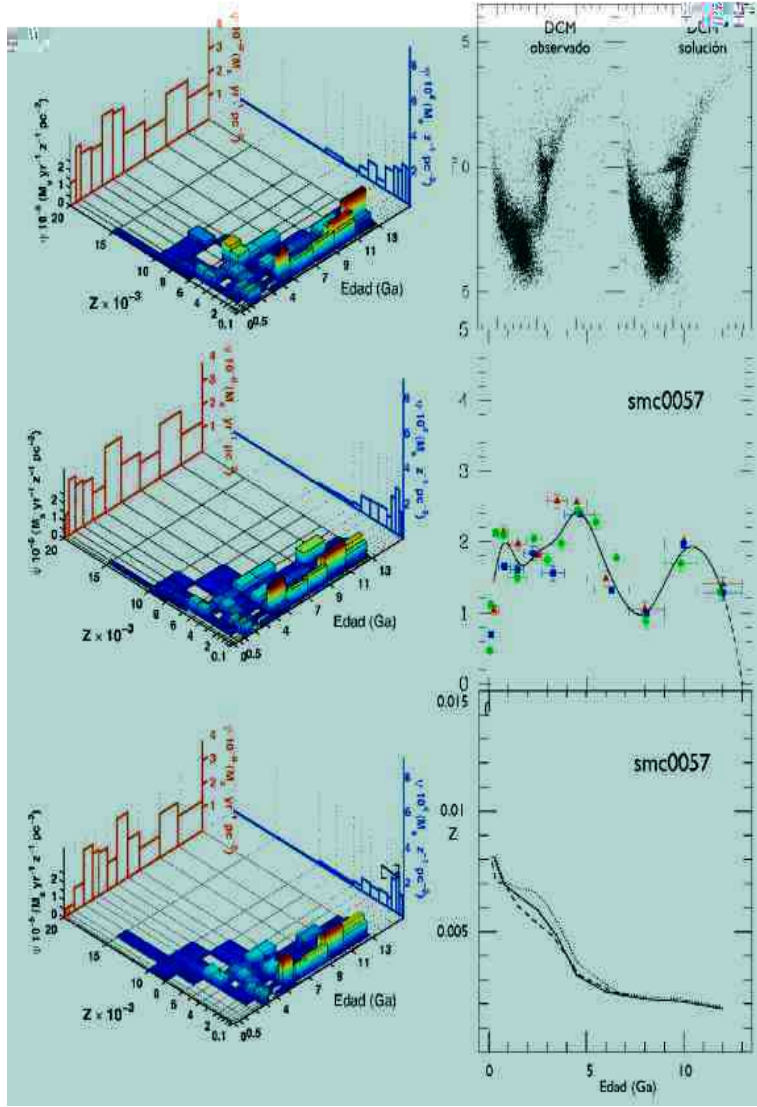}
}
\caption{Left panel:  three-dimensional representation of the solution 
for the SFH of the labelled field. The x-axis shows the age of the stars (in Gyr), 
the y-axis shows the metallicity of the stars, and the z-axis shows $\psi(t)$, 
in units of solar masses per year, metallicity interval, and area. $\psi(t,z)$ 
in each age-interval is given by the height of the bar emerging from the xy plane. 
Right panel: observed and solution CMDs (above),  $\psi(t)$ solutions for the 
three age binnings (middle) and  corresponding age-metallicity relations (bottom).  
Each of the individual $\psi(t)$ solutions were corrected by the systematic errors 
discussed in Sec.~\ref{mock} and are represented by a different symbol and color: 
red triangles are for age-1,  blue squares are for age-2, and green circles
are for age-3.   Each $\psi(t)$ point carries its vertical error bar that
is the formal error from IAC-pop, calculated  as the dispersion of 20
solutions with   $\chi^{2}_{\nu}$=$\chi^{2}_{\nu,min}$+1, where 
$\chi^{2}_{\nu,min}$ is that of the solution shown in the figure (see
Aparicio \& Hidalgo 2009).   Horizontal tracks are not error bars but show
the age interval associated to each point.  The results for the three age-binning sets were combined 
by fitting a cubic spline.   We do not have a constraint on
$\psi(t)$  at 13 Gyr old and so the end point of our spline fit is
arbitrary. Choosing zero for the end point  gives good agreement  between
the integrated SFH under our spline fit and those of our measured SFHs for
the three age binnings.   }
\label{figure_3D}
\end{center}
\end{figure*}

\begin{figure}
\plotone{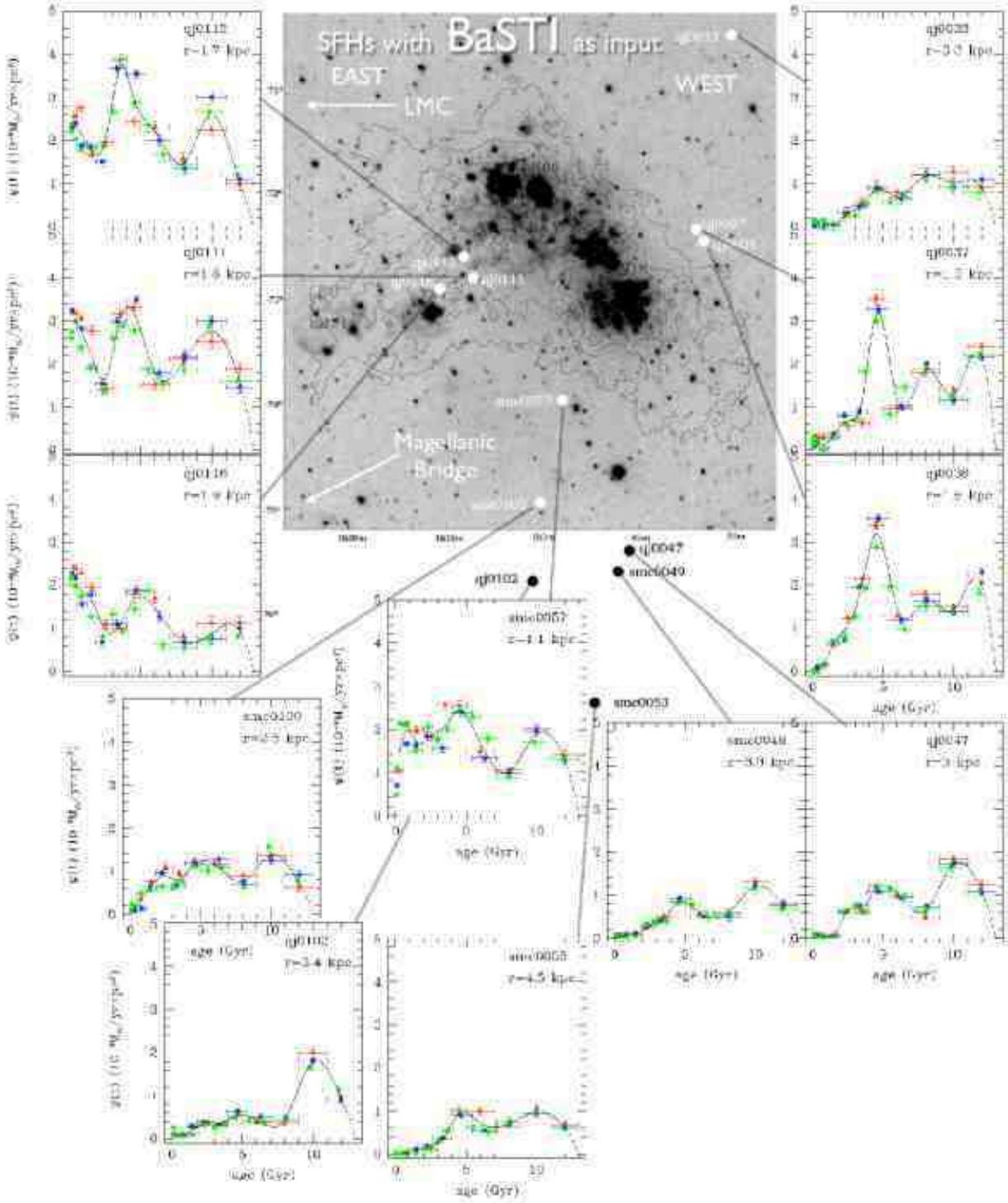}
\caption{The derived SFHs of our SMC fields. BaSTI stellar evolution library was used as input of  IAC-star.
  Each solution shows the SFH obtained for 3 three age binning schemes from table~\ref{age_intervals} 
(red triangles: age-1, blue squares age-2, and green circles age-3). Each point carries its vertical error bar that is the formal error 
from IAC-pop, calculated  as the dispersion of 20 solutions with  
$\chi^{2}_{\nu}$=$\chi^{2}_{\nu,min}$+1.  
 Horizontal tracks are not error bars, but show the age interval associated to each point. 
 The solid line shows the results of a cubic spline fit to the results.
We do not have a constraint on the $\psi(t)$ 
 at 13 Gyr old and the end point of our spline fit was chosen to be zero arbitrarily (dashed lines between 12 and 13 Gyr ago
 in the spline fit).  See text for details.
\label{SFH_I}}
\end{figure}

\begin{figure}
\plotone{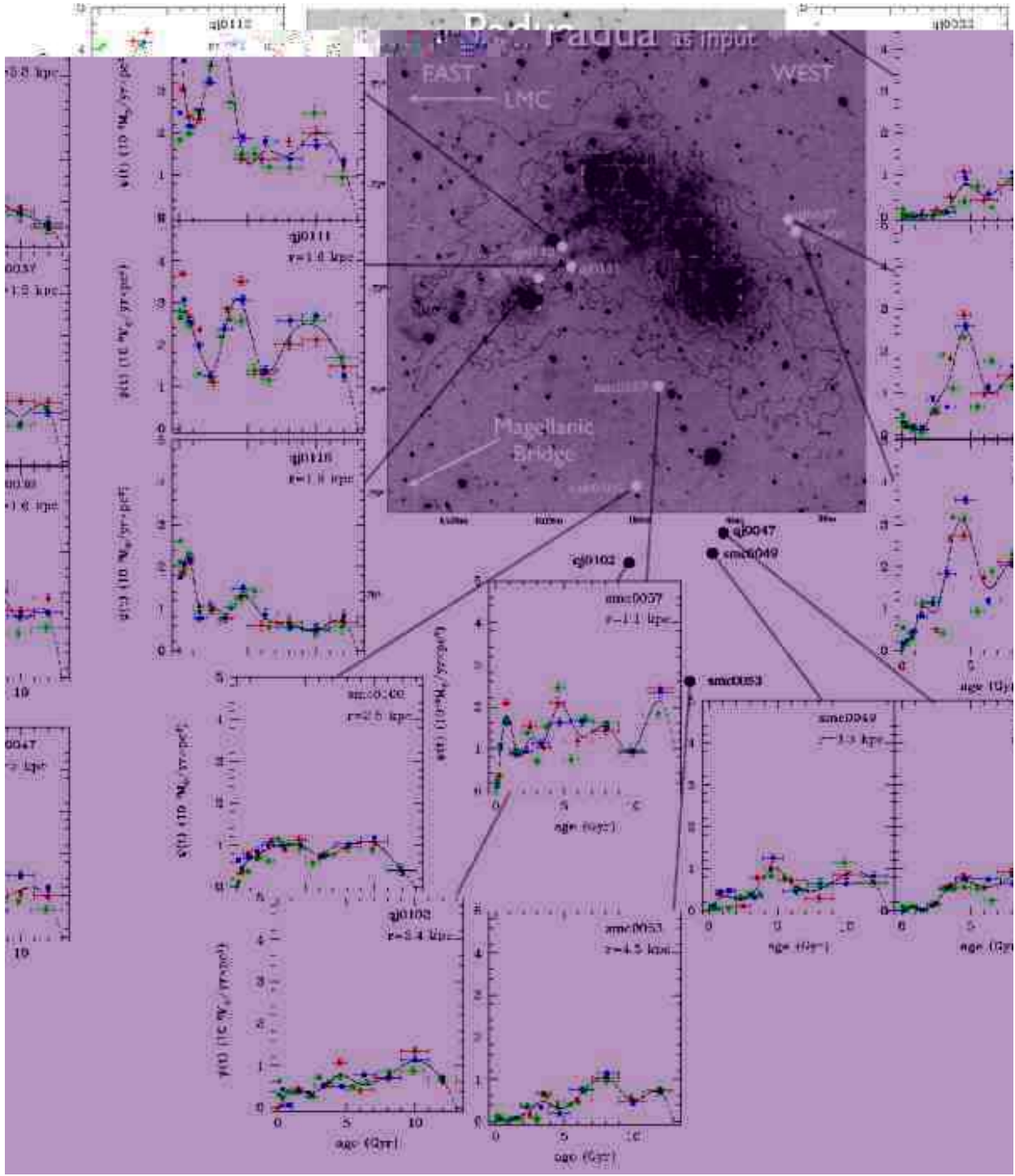}
\caption{Same as figure~\ref{SFH_I} but using Padua stellar evolution library  as input of  IAC-star. Note that the main
characteristics seen in figure~\ref{SFH_I} are impervious to the change of stellar evolution library. \label{noeplot_final_bertelli}}
\end{figure}

\begin{figure*}
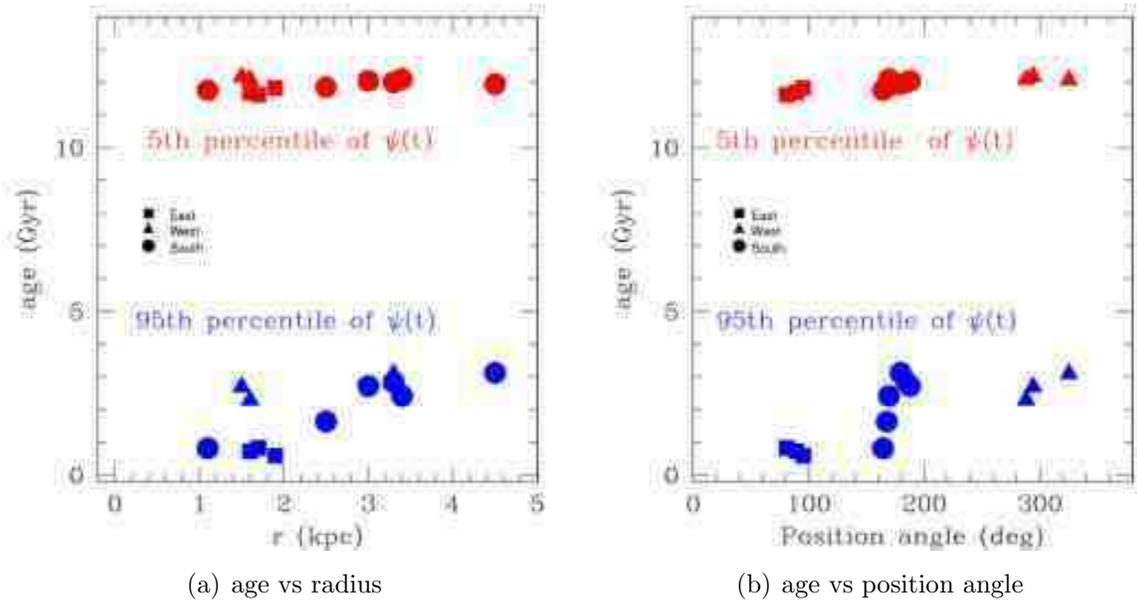

\begin{center}\subfigure[age vs radius]
	{
		\label{percentil_d}
\includegraphics[width=7.5cm,height=7.5cm]{percentil_d.eps}
}
\subfigure[age vs position angle]
	{
		\label{percentil_pa}
\includegraphics[width=7.5cm,height=7.5cm]{percentil_pa.eps}
         }
\caption{The age at the 5th and at the 95th percentiles of $\psi(t)$
 for each of our SMC fields are  represented as a function of radius, position angle, and
age. See text for details.}
\label{percentiles}
\end{center}
\end{figure*}

\begin{figure*}
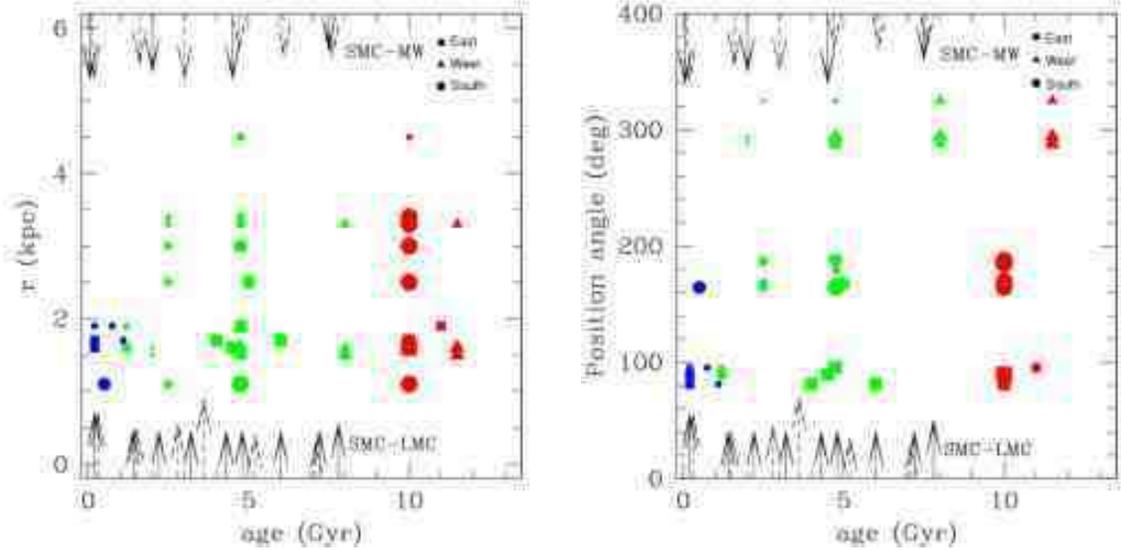

\begin{center}\subfigure[Intensity of $\psi(t)$ enhancements (age vs. radius).]
      {
	      \label{enhancenew_d}
\includegraphics[width=7.5cm,height=7.5cm]{enhancenew_d.eps}
}
\subfigure[Intensity of $\psi(t)$ enhancements (age vs. position angle).]
      {
	      \label{enhancenew_pa}
\includegraphics[width=7.5cm,height=7.5cm]{enhancenew_pa.eps}
 	 }
\caption{Intensity of the $\psi(t)$ enhancements together with pericenter passages of the SMC.
 We fitted a gaussian function to the elevations in figure~\ref{SFH_I}.
The size of the symbols depends on the intensity
of the enhancement.
   The bottom  arrows indicate the pericenter passages with the LMC while the top arrows 
     show the encounters with the Milky Way 
     (solid-lined arrow represent data  from Kallivayalil et al. 2006 and dashed-lined ones are data obtained 
     from Bekki \& Chiba 2005). The size of the arrows represent the intensity of the encounter.
     Note that some enhancements are hidden behind larger ones.
      See text for details.}
\label{enhancements}
\end{center}
\end{figure*}

\begin{figure*}
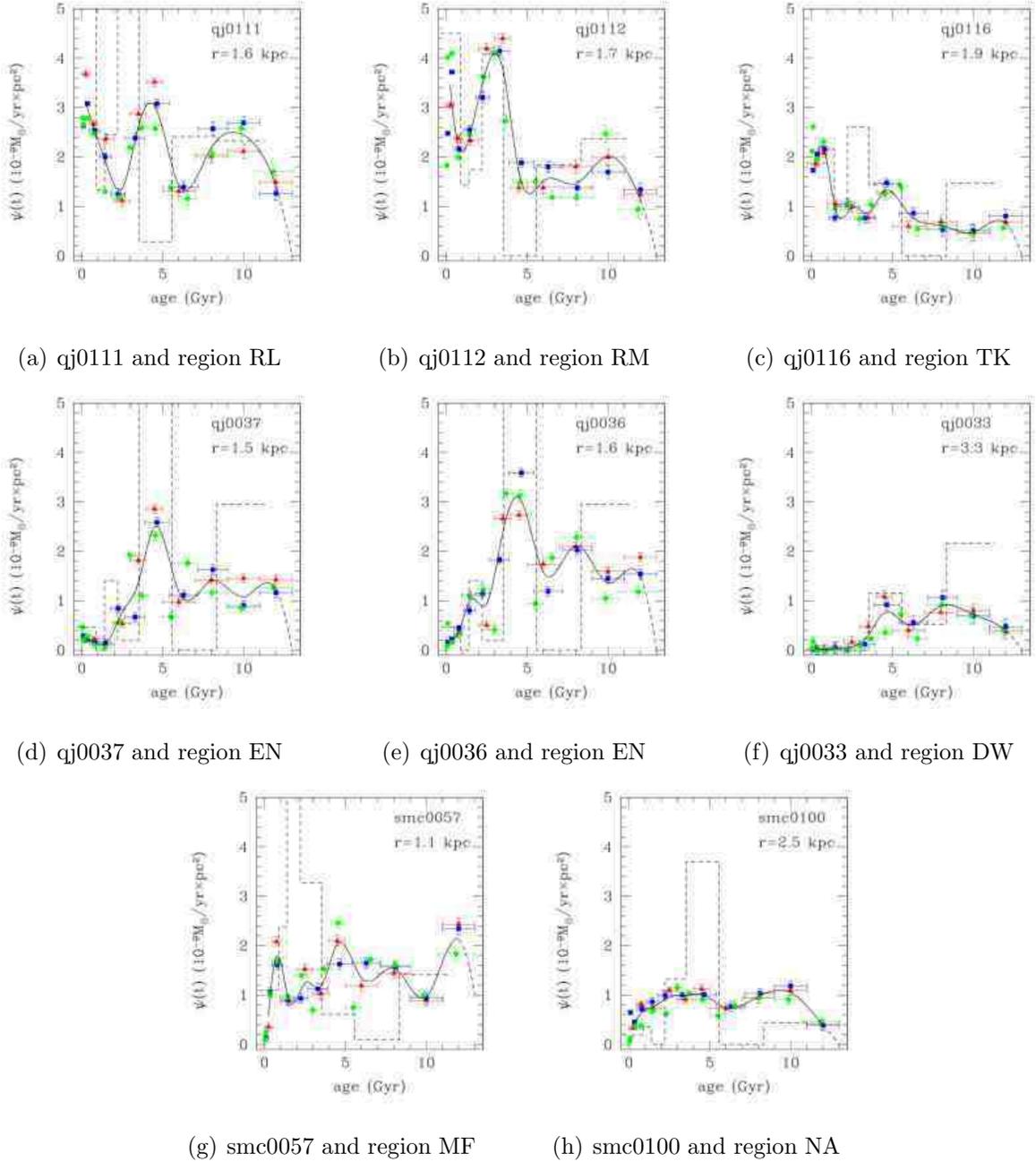

\begin{center}
\subfigure[qj0111 and region RL]
	{
			\label{HZ04_qj0111}
\includegraphics[width=5cm,height=5cm]{qj0111final_bertelli_RL_HZ.eps} 
}
\subfigure[qj0112 and region RM]
	{
		\label{HZ04_qj0112}
\includegraphics[width=5cm,height=5cm]{qj0112final_bertelli_RM_HZ.eps} 
}
\subfigure[qj0116 and region TK]
	{
		\label{HZ04_qj0116}
\includegraphics[width=5cm,height=5cm]{qj0116final_bertelli_TK_HZ.eps} 
}
\subfigure[qj0037 and region EN]
	{
		\label{HZ04_qj0037}
\includegraphics[width=5cm,height=5cm]{qj0037final_bertelli_EN_HZ.eps} 
}
\subfigure[qj0036 and region EN]
	{
		\label{HZ04_qj0036}
\includegraphics[width=5cm,height=5cm]{qj0036final_bertelli_EN_HZ.eps} 
}
\subfigure[qj0033 and region DW]
	{
		\label{HZ04_qj0033}
\includegraphics[width=5cm,height=5cm]{qj0033final_bertelli_DW_HZ.eps} 
}
\subfigure[smc0057 and region MF]
	{
		\label{HZ04_smc0057_BaSTI}
\includegraphics[width=5cm,height=5cm]{smc0057final_bertelli_MF_HZ.eps} 
}
\subfigure[smc0100 and region NA]
	{
		\label{HZ04_smc0100}
\includegraphics[width=5cm,height=5cm]{smc0100final_bertelli_NA_HZ.eps} 
}
\caption{Comparison of the SFHs obtained in this work using the Padua stellar evolution library as input  in IAC-star
 (see figure~\ref{noeplot_final_bertelli}) 
and the ones obtained by HZ04 (dashed lines) for the overlapping fields.   
Their EN region covers a larger area, including both fields qj0036 and qj0037. See text for details.}
\label{HZ04}
\end{center}
\end{figure*}

\newpage

\begin{figure*}
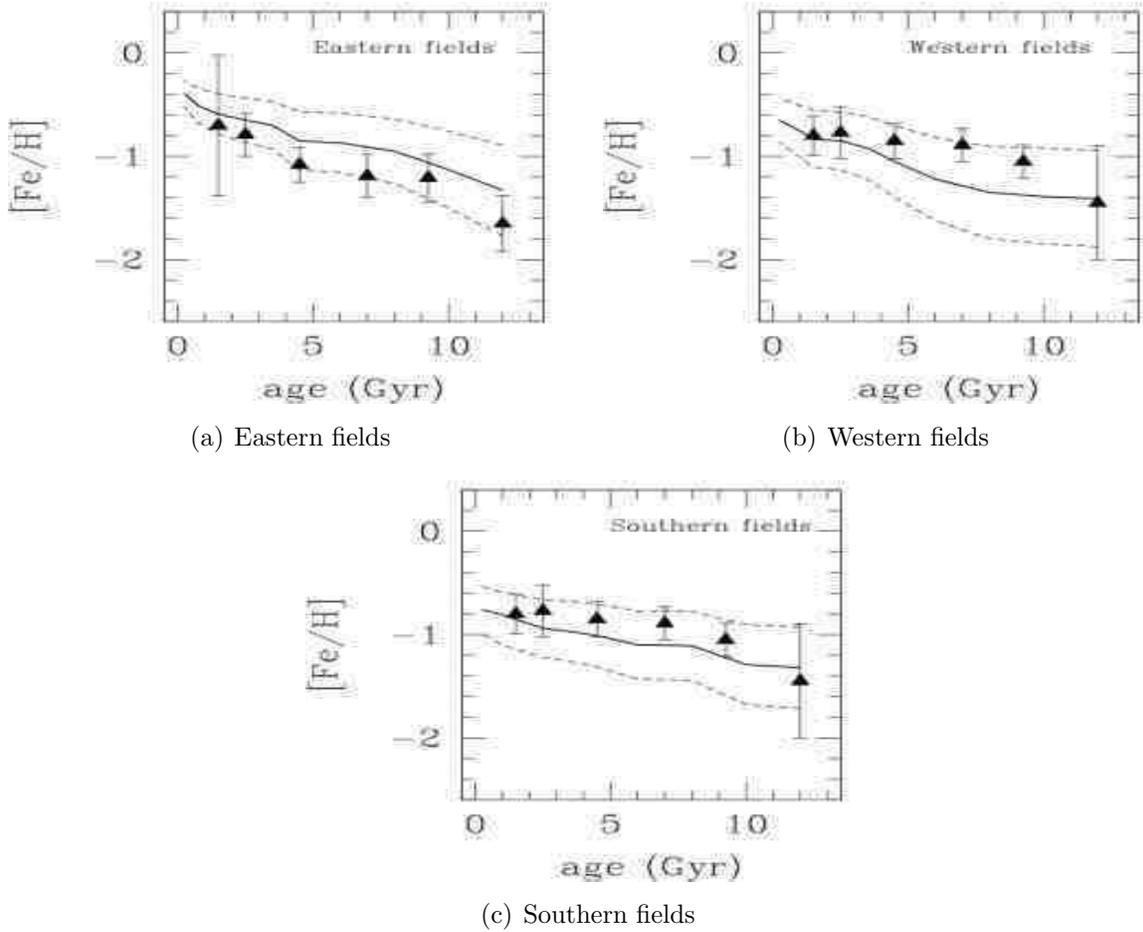

\begin{center}
\subfigure[Eastern fields]
	{
		\label{ceh_east}
\includegraphics[width=7.5cm,height=5.5cm]{ceh_mediana_east.eps} 
}
\subfigure[Western fields]
	{
		\label{ceh_west}
\includegraphics[width=7.5cm,height=5.5cm]{ceh_mediana_west.eps} 
}
\subfigure[Southern fields]
	{
		\label{ceh_south}
\includegraphics[width=7.5cm,height=5.5cm]{ceh_mediana_south.eps} 
}
\caption{The averaged age-metallicity relations ([Fe/H] as a function of age) for the eastern, western, and southern fields are
shown with a solid line. The dotted lines represent the $\pm$1$\sigma$ level. 
  The triangles represent  the age-metallicity relations   
  found by Carrera et al. (2008b) (see their table 6).}
\label{ceh}
\end{center}
\end{figure*}



\end{document}